\documentclass[aps,pra,reprint,superscriptaddress, footinbib]{revtex4-1}
\usepackage{graphicx}
\usepackage{amsmath}
\usepackage{amssymb}
\usepackage{amsbsy}
\usepackage{siunitx}

\begin{document}
\title{Probing electron-hole components of subgap states in Coulomb blockaded Majorana islands}

% autoriai %
\author{Esben Bork Hansen}
\affiliation{Center for Quantum Devices and Station Q Copenhagen, Niels Bohr Institute, University of Copenhagen, Copenhagen 2100, Denmark}
\author{Jeroen Danon}
\affiliation{Center for Quantum Spintronics, Department of Physics, Norwegian University of Science and Technology, NO-7491 Trondheim, Norway}
\author{Karsten Flensberg}
\affiliation{Center for Quantum Devices and Station Q Copenhagen, Niels Bohr Institute, University of Copenhagen, Copenhagen 2100, Denmark}

\date{\today}
\begin{abstract}
Recent tunneling spectroscopy experiments in semiconducting nanowires with proximity-induced superconductivity have reported robust zero-bias conductance peaks. Such a feature can be compatible with the existence of topological Majorana bound states (MBSs) and with a trivial Andreev bound state (ABS) near zero energy. Here, we argue that additional information, that can distinguish between the two cases, can be extracted from Coulomb-blockade experiments of Majorana islands. The key is the ratio of peak heights of consecutive conductance peaks give information about the electron and hole components of the lowest-energy subgap state. In the MBS case, this ratio goes to one half for long wires, while for short wires with finite MBS overlap it oscillates a function of Zeeman energy with the same period as the MBS energy splitting. We explain how the additional information might help to distinguish a trivial ABS at zero energy from a true MBS and show case examples.
\end{abstract}
%-----------------------------------------------------------%

\maketitle

A semiconductor-superconductor hybrid nanowire system can exhibit a topological p-wave superconducting phase due to the interplay between Rashba spin-orbit coupling, induced s-wave superconductivity, and an appropriately applied Zeeman field~\cite{Lutchyn2010,Oreg2010}. The p-wave superconductor is of great interest because it can host Majorana bound states (MBS)~\cite{Kitaev2001}, that may serve as elementary building blocks of a topologically protected quantum computer~\cite{Kitaev2003}.

In the last decade, the hunt for MBSs has led to an extensive study of these so-called Majorana nanowires, along with a growing list of theoretical predicted features of MBSs in these systems. The list includes an exponential suppression of MBS energy with the length of the wire~\cite{Kitaev2001}, a $4\pi$-periodic Josephson effect~\cite{Kitaev2001}, a ${\rm 2e^2}/h$-quantized zero-bias conductance peak~\cite{Law2009,Sengupta2001,Flensberg2010}, and non-abelian braiding statistics~\cite{Nayak2008}. Since the first observations of a zero-bias peak on the background of a soft superconducting gap~\cite{Mourik2012}, advancements in material growth have enhanced the quality and resolution of experiments to a point where the more detailed features of the possible MBSs can be subjected to further experimental tests. The clean interface between Al and InAs in epitaxial nanowires has been shown to induce a hard superconducting gap in the nanowire, close to the gap of Al~\cite{Chang2015}, which in turn enabled the observation of an exponential suppression of the oscillations of the lowest bound-state energy with increasing wire length in Coulomb-blockaded Majorana islands (CBMI)~\cite{Albrecht2016}, as well as, more recently, a quantized zero-bias conductance of $2{\rm e}^2/h$~\cite{Nichele2017,Zhang2017}. 

However, persistent zero-bias peaks in conductance measurements are not conclusive evidence for the existence of MBSs since other (topologically trivial) phenomena might give rise to zero-bias peaks as well. Trivial Andreev bound states (ABSs) with conductance features resembling MBSs might arise due to disorder~\cite{Liu2012}, smooth confinement~\cite{Kells2012}, and/or strongly coupled non-superconducting quantum dots at the ends of the nanowire~\cite{Chiu2017}. Braiding experiments would give a conclusive answer to whether the states associated with the observed zero-bias peak are topological or trivial in nature, but since these are still outside the reach of current experiments, transport spectroscopy is currently among the best techniques for obtaining details of the quantum states of Majorana nanowires. It is therefore of great interest to the field to extract additional information about the state behind the zero-bias peak from the currently accessible transport spectroscopy measurements. 

Experiments with CBMIs have shown consistent behavior in tunnel conductance measurements over many Coulomb peaks~\cite{Albrecht2016}, indicating that transport happens through the same state and that the state is, to a large degree, unperturbed by the change in gate voltage. So far, analyses of the zero-bias conductance in these setups have mainly focused on the oscillations and intensity of individual peaks as a function of system parameters~\cite{Albrecht2016,Heck2016,Sarma2012}.

\begin{figure}
	\centering
	\includegraphics[width=0.95\linewidth]{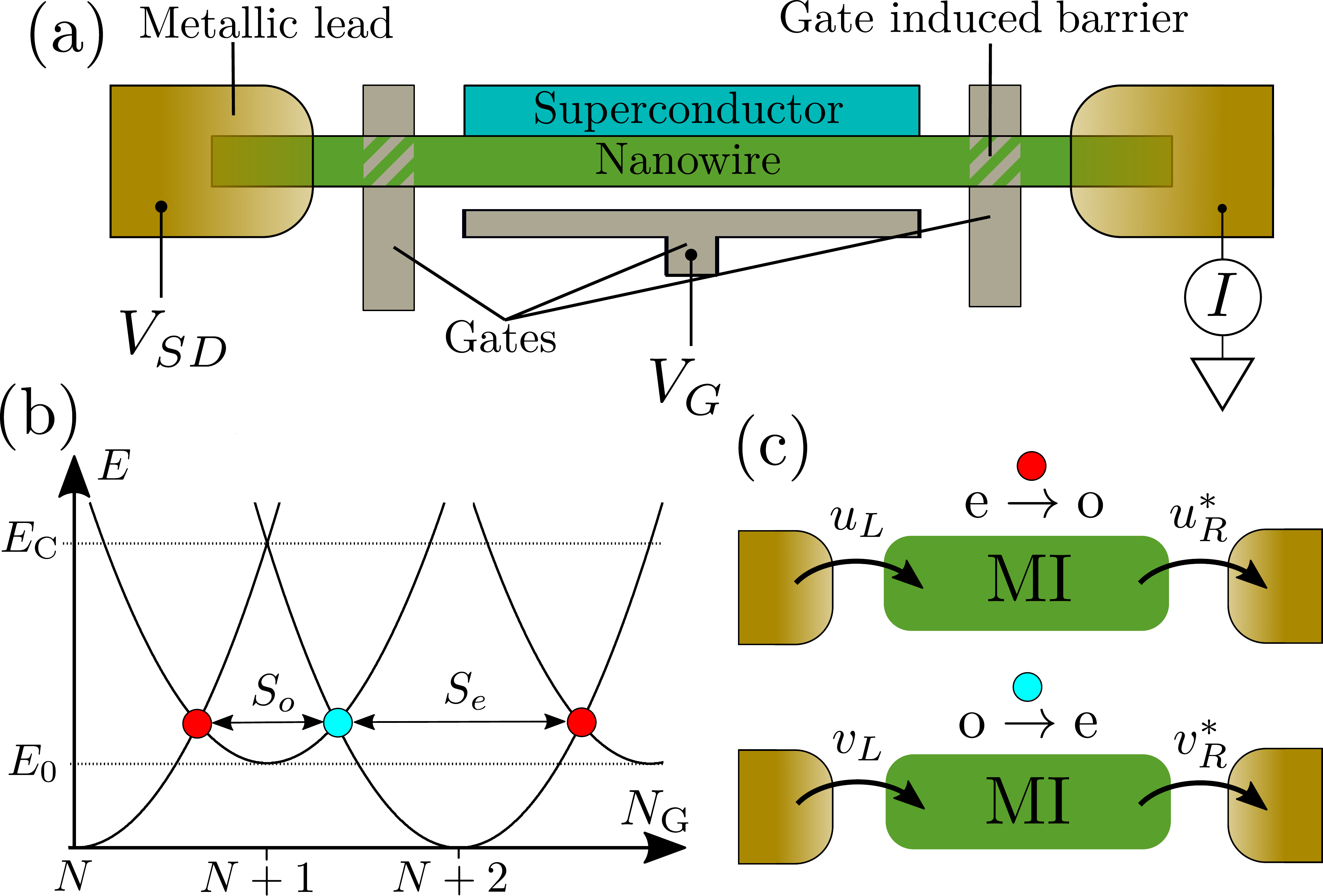}
	\caption{[color online] (a) Schematic of a Coulomb blockaded Majorana island. A section of the semiconducting nanowire is partly covered by a superconductor, inducing superconductivity into the nanowire. At the ends of the superconducting region, gates are used to induce tunneling barriers to metallic leads. A gate with voltage $V_G$ is placed on the side of the nanowire to control the number of charges on the island. (b) Energy spectrum of the CBMI, given in Eq.~(\ref{eq.energy_spectum}), as a function of the dimensionless gate-induced charge $N_G$. (c) At even-to-odd (even-to-odd) charge-state degeneracies, indicated in (b), the sequential transport is governed by the value of the norm of the coherence factor $|u|^2$ ($|v|^2$) at the end of the wire.}
	\label{fig:fig1}
\end{figure}

In this paper, we will discuss how zero bias conductance (ZBC) measurements on CBMIs [see Fig.~\ref{fig:fig1}(a)] can give information about the electron and hole components of the system's lowest-energy state, which in turn might help to discern whether this state is a MBS or a trivial ABS. Assuming that the lowest energy-state is well separated from higher excited states on the scale of temperature and tunnel coupling, the ZBC at even-odd (odd-even) charge degeneracies will in the sequential tunneling regime be proportional to the electron (hole) component of the lowest energy state, see Fig.~\ref{fig:fig1}(b,c). The ratio of consecutive ZBC peaks gives a direct measure of the ratio of the electron and hole components of the state, which can be compared with theoretical predictions. For a MBS, we find that this ratio will follow a similar beating pattern as that of the energy splitting. This additional piece of information might serve to discern a trivial ABS from a true MBS. We show how this applies to a case example, similar to setups considered in~\cite{Chiu2017} and~\cite{Moore2017}, where a Majorana nanowire with a non-superconducting region at the end hosts a trivial Andreev bound state that gives rise to a zero bias peak similar to what is seen in the experiment reported in~\cite{Deng2016}, and we find that in this case a trivial state can be distinguished from a topological MBS.

MBSs are localized exponentially and the wavefunction overlap of a pair of MBSs is exponentially suppressed by the distance between them. When the separation of MBSs is several times larger than the Majorana localization length, the MBSs are to a good approximation completely decoupled from each other and consist of an equal superposition of an electron and a hole at zero energy. In shorter nanowires, where the overlap between MBSs is non-negligible, the MBSs acquire an energy splitting, which oscillates as a function of magnetic field. The electron and hole components ($u$ and $v$) of overlapping MBSs are in general unequal, and the ratio $|u|^2/(|v|^2+|u|^2)$ oscillates as a function of magnetic field, with the same period as the energy splitting, but shifted such that the ratio differs from $1/2$ when the splitting is zero and is $1/2$ when the splitting is maximal, see Fig.~\ref{fig:fig3}(d).

\textit{Model:} 
We consider the setup of a Coulomb-blockaded Majorana island as depicted in Fig.~\ref{fig:fig1}(a). It is assumed that the subgap eigenstate of the MI with lowest energy $E_0$ is well separated from higher eigenstates on the scale of temperature and tunnel coupling to the leads. In the sequential-tunneling regime, the dominating contribution to conductance at zero bias is by transport through this state and we will therefore include only this state in our transport model. We label the configurations of the CBMI by the number of charges $N$ and the occupation of the state at $E_0$, $N_0=\{0,1\}$, which can only be odd (even) when $N_0 = 1 (0)$. The spectrum, shown in Fig.~\ref{fig:fig1}(b), is given by
\begin{equation}
E(N,N_G,N_0) = E_C(N-N_G)^2 + N_0 E_0,\label{eq.energy_spectum}
\end{equation}
where the first term is the electrostatic energy due to Coulomb interaction, with $N_G$ being the dimensionless gate-induced charge proportional to the gate voltage $V_G$. At zero bias, conductance peaks appear when two charge states are degenerate. The distance between these peaks is labeled $S_{\rm o}$ or $S_{\rm e}$ corresponding to odd or even ground states, see Fig~\ref{fig:fig1}(b). 

To describe the electron transport, we employ a set of master equations with transition rates to lowest order in the tunneling Hamiltonian between the MI and the leads. In a non-superconducting Coulomb-blockaded island, a state is always filled (emptied) when an electron enters (exits) the island, but in the presence of superconductivity a state can also be emptied (filled) when an electron enters (exits) the island by the simultaneous creation (annihilation) of a Cooper pair. The creation operator that creates an electron with spin $\sigma$ at position $x$ on the MI is thus a superposition of quasiparticle creation and annihilation operators,
\begin{equation}\label{Psixdef}
\psi_\sigma^\dagger(x) =\sum_{n} \left(u_{n\sigma}(x)\gamma_{n}^\dagger+ v^*_{n\sigma}(x)\gamma_{n}S^\dagger\right),
\end{equation}
where the sum is over all eigenstates of the MI and the operator $S^\dagger$ creates a Cooper pair. The coherence factors $u$ and $v$ depend on the microscopic model of the MI which we will return to later. Since the tunnel coupling involves actual electrons, the tunneling Hamiltonian can be written in terms of quasiparticles using Eq.~\eqref{Psixdef}. Projecting the tunneling Hamiltonian onto the lowest bound state, with creation operator $\gamma_0^\dagger$, we write
\begin{equation}\label{HT}
H_T = \sum_{\sigma\alpha=L,R} t_\alpha \left(u_{0\sigma,\alpha}\gamma_{0}^\dagger +v_{0\sigma,\alpha}^*\gamma_{0}^{{}} S^\dagger\right)c_{\sigma\alpha\nu}^{{}} + \mathrm{h.c.},
\end{equation}
where $c_{\sigma\alpha\nu}^\dagger$ is the is the electronic creation operator of electrons in lead $\alpha$ with orbital index $\nu$. We assume a constant density of states in the metallic leads and energy-independent tunnel couplings. The coherence factors $u_{0\sigma,\alpha}$ and $v_{0\sigma,\alpha}$ correspond to the coherence factors at the left ($x = 0$) and right ($x = L$) end of the island for $\alpha= L$
and $R$, respectively. For further details on the master-equation model and the calculation, see e.g.~the supplementary material of~\cite{Albrecht2017} \cite{note1}.

The zero-bias conductance can be calculated analytically in the vicinity of ground-state degeneracies by solving the master equation under the approximation that only the two degenerate states contribute to transport:
\begin{equation}
G_{{\rm e} \to {\rm o}} = \frac{e^2}{h}\frac{\gamma_L\gamma_R|u_Lu_R|^2}{4(\gamma_L|u_L|^2+\gamma_R|u_R|^2)}\frac{\beta}{\cosh^2(\frac{\beta}{2}(\delta E^{(N)}_{\text{el}}+E_0))}, \label{eq.G_eo}
\end{equation}
for an even-to-odd degeneracy [see Fig.~\ref{fig:fig1}], where $\gamma_{R/L}$ is a dimensionless parameter characterizing the coupling to the right/left lead and $\beta$ is the inverse temperature. We introduced the electrostatic energy difference between charge states $N+1$ and $N$, $\delta E_{\rm el}^{(N)} = E_{\rm el}(N+1) - E_{\rm el}(N)$, and the spin sum of the coherence factor $|u_{\alpha}|^2 = \sum_\sigma |u_{\sigma\alpha}|^2$. The conductance at the odd-to-even degeneracy $G_{\text{o}\rightarrow \text{e}}$ is the same as Eq.~(\ref{eq.G_eo}), but with $u_\alpha \leftrightarrow v_\alpha$ and $E_0\leftrightarrow-E_0$ interchanged.

Assuming that the coherence factors $u_\alpha$ and $v_\alpha$ do not change with $V_G$ and that the tunnel couplings at both ends are equal, $\gamma_R = \gamma_L$, we obtain a measure of the coherence factors of the bound state
\begin{equation}
\Lambda=\frac{G_{\text{e}\rightarrow \text{o}}}{G_{\text{e}\rightarrow \text{o}} + G_{\text{o}\rightarrow \text{e}}} = \frac{\frac{|u_L|^2|u_R|^2}{|u_L|^2 + |u_R|^2}}{\frac{|u_L|^2|u_R|^2}{|u_L|^2 + |u_R|^2} + \frac{|v_L|^2|v_R|^2}{|v_L|^2 + |v_R|^2}},\label{eq.ratio}
\end{equation}
where $G_{\rm e\rightarrow o}$ and $G_{\rm o\rightarrow e}$ are (consecutive) conductance peaks, at $\delta E_{\rm el}^{(N)} + E_0=0$ and $\delta E_{\rm el}^{(N+1)} - E_0=0$, respectively.
$\Lambda$ thus takes values between 0 and 1, where 1(0) correspond to the state being purely electron(hole)-like at the ends of the wire.
This result holds in general, as long as the lowest-energy state is well separated from higher-energy states and sequential tunneling processes are dominating the transport. This ratio of the measured conductances can then be compared with theoretical predictions, which we will discuss in the next section.

The constant interaction model in Eq.~(\ref{eq.energy_spectum}) assumes that the single-particle spectrum does not depend on the number of particles on the island. This assumption can be justified for islands where the level spacing is much smaller than the charging energy \cite{Aleiner2002}, a condition which is definitely fulfilled for the metallic island considered here. Moreover, it is important for our analysis that the subgap state (and hence the coherence factors) relevant for the transport does not change with electron number, such that many consecutive peaks heights can be described by the same coherence factors. In other words, the chemical potential (set by the superconductor) in the single-particle Hamiltonian should be constant over large number of Coulomb-blockade peaks. 

We calculate the change in chemical potential due to changing the number of electrons in the superconducting shell by $\delta N$ by
\begin{equation}
\delta \mu = \frac{\delta N}{\mathcal{V}d(\epsilon_F)},
\end{equation}
where $d(\epsilon_F)\approx23$~\si{\nano\meter^{-3} \electronvolt^{-1}} is the density of states at the Fermi energy in the Al superconducting shell~\cite{Court2008} and the volume of the shell $\mathcal{V} \sim 10^6$ \si{\nano\meter^3}. The volume of the shell is estimated for a nanowire of length $1$~\si{\micro\meter}, diameter $100$~\si{\nano\meter}, Al on two facets, and shell of thickness $10$~\si{\nano\meter}. Assuming that the density of states of the shell is constant in the considered energy range, we estimate the change in chemical potential is of the order $\delta\mu\sim 50$~\si{\pico\electronvolt} per electron. We calculate the change in the average number of electrons in the semiconductor nanowire $\delta Q$ due to a change in Zeeman field of $1$~\si{\milli\electronvolt} with chemical potential kept constant by numerically diagonalizing the Hamiltonian given below and find that $\delta Q\sim 5$ e. Redistributing this small number of electrons from the superconducting shell to the wire leads to a change in chemical potential that is much smaller than all relevant energy scales in the system and can assumed to be constant. 

We model the non-interacting Majorana nanowire by a single channel BdG Hamiltonian~\cite{Lutchyn2010,Oreg2010}
$H = \frac{1}{2}\int \Psi^\dag(x)\mathcal{H}\Psi(x)dx,$
using the Nambu spinor $\Psi^\dag = (\psi^\dag_\uparrow, \psi^\dag_\downarrow, \psi_\downarrow, -\psi_\uparrow)$ and single particle Hamiltonian
\begin{align}
\mathcal{H} &= (-\frac{\hbar^2}{2m^*}\partial_x^2 - \mu - i\alpha_{\rm R}\partial_x\sigma_y)\tau_z + V_{\mathrm{Z}}\sigma_x + \Delta\tau_x,\label{eq:hamiltonian}
\end{align}
where $m^*$ is the effective mass of the electrons, $\mu$ is the chemical potential, $\alpha_{\mathrm{R}}$ is the Rashba spin-orbit coupling strength, $V_{\mathrm{Z}}=\frac{1}{2}g\mu_{\text{B}}B$ is the Zeeman energy due to the magnetic field $B$ along the nanowire with the Land\'{e} g-factor $g$ and Bohr magneton $\mu_{\text{B}}$, and $\Delta$ is the induced superconducting gap. The Pauli matrices $\pmb{\sigma}$ and $\pmb{\tau}$ act on spin and particle-hole space, respectively.

The eigenenergies $\epsilon_n$ and the electron and hole components $u_{n\sigma}$ and $v_{n\sigma}$, are obtained by numerically diagonalizing a discretized version of the Hamiltonian on a chain of $N=100$ sites with length $L=1$~\si{\micro\meter}. Other parameters are the effective mass of the electrons in the nanowire $m^*=0.026m_{\rm e}$, spin-orbit coupling strength $\alpha_{\mathrm{R}} = 0.3$~\si{\electronvolt\angstrom}, chemical potential $\mu=0$~\si{\micro\electronvolt}, and superconducting gap $\Delta=140$~\si{\micro\electronvolt}.

\begin{figure}
	\centering
	\includegraphics[width=1\linewidth]{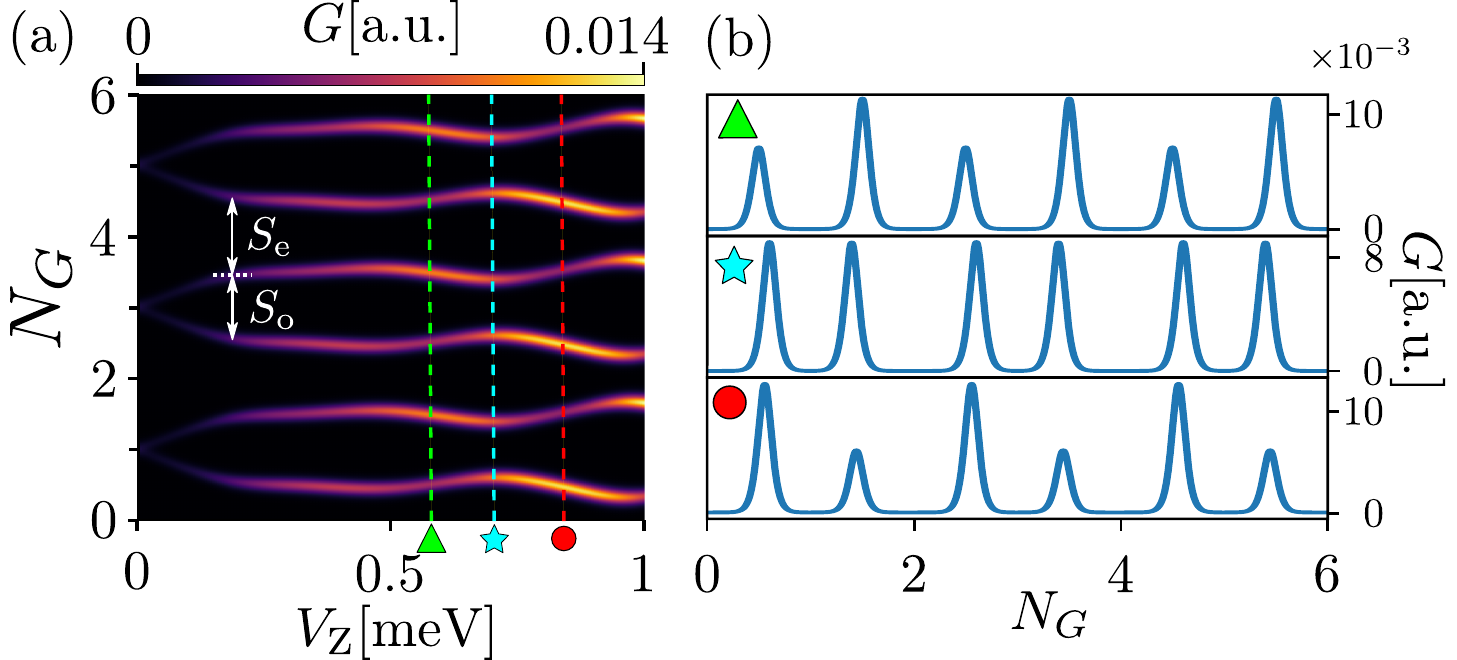}
	\caption{[color online] (a) Zero-bias conductance as a function of Zeeman field $V_{\rm Z}$ and gate-induced charge $N_G$, calculated from Eq.~(\ref{eq.G_eo}), with $T=150$~mK and $\gamma_{R/L}=1$. Colored dashed lines indicate where the cuts in (b) are taken. (b) Cuts along the $N_G$-axis of (a) at $V_{\rm Z}= 570$~\si{\micro\electronvolt} (green triangle), 685~\si{\micro\electronvolt} (cyan star), and 820 \si{\micro\electronvolt} (red circle).}
	\label{fig:fig2}
\end{figure}

In Fig.~\ref{fig:fig2}, we show a simulated example how measured data of a CBMI might look for the parameters given above (comparable to those reported in experiments). Zero-bias conductance is calculated using Eq.~(\ref{eq.G_eo}), where the energy $E_0$ and the coherence factors at the end of the nanowire $u_\alpha$ and $v_\alpha$ are obtained numerically. Other parameters are the temperature $T=150$~\si{\milli\kelvin}, coupling to leads $\gamma_L=\gamma_R=1$, and charging energy $E_C=150$~\si{\micro\electronvolt}. In a measured data set like this, the parameter $\Lambda$, as defined in Eq.~(\ref{eq.ratio}), can then be accessed directly by considering the relative height of consecutive conductance peaks, as shown in \ref{fig:fig2}(b).

In Fig.~\ref{fig:fig3}(d), we plot $S_{\rm e} - S_{\rm o}$ (red) and $S_{\rm o} - S_{\rm e}$ (blue), which correspond to the energy $\pm E_0$ and plot it along with the ratio $\Lambda$. The oscillating behavior of $\Lambda$ after the topological phase transition is a generic feature of MBSs. The oscillations follow the same period as the splitting between $\pm E_0$, but shifted so that $|\Lambda|$ is maximal when $E_0$ crosses zero. That is, the MBS is more electron/hole-like when it is at zero energy and half electron, half hole when the energy splitting is maximal. This behavior is generic in the sense that $\Lambda$ is correlated to the oscillations of $E_0$, such that if parameters are changed, $\Lambda$ will change accordingly to the change of $E_0$~\cite{Ben-Shach2015, Domínguez2016}.

\begin{figure}
	\centering
	\includegraphics[width=1\linewidth]{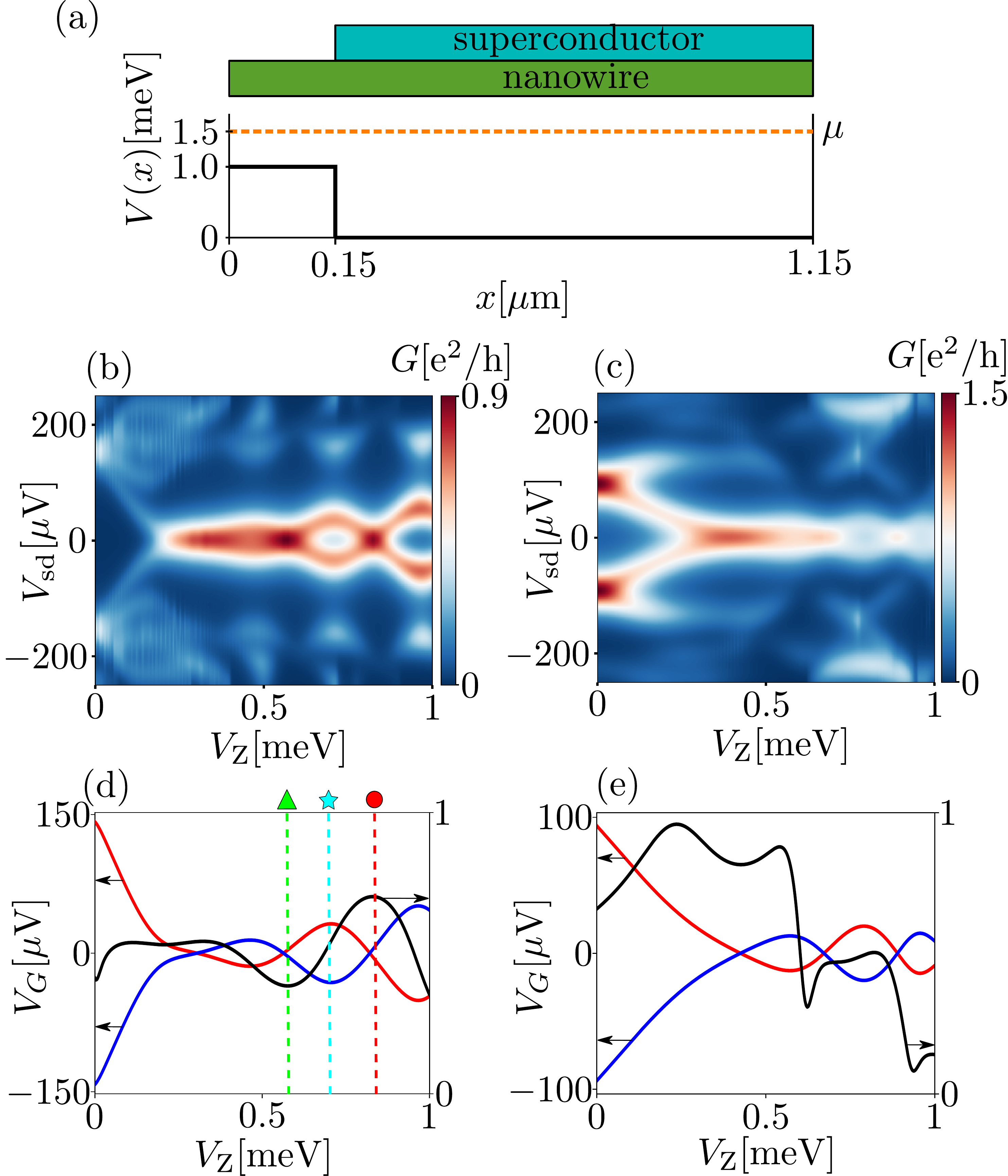}
	\caption{[color online] (a) Schematic of a case-example setup (where a section of the nanowire is not covered by the superconductor) and the local potential $V_(x)$ in the nanowire. (b,d) are for the setup described in the previous section, while (c,e) are for this case example. (b,c) Differential conductance for tunneling into the left end of a grounded nanowire calculated using the method described in~\cite{Hansen2016}. (d,e) $S_{\rm e}-S_{\rm o}$ (red) and $S_{\rm o}-S_{\rm e}$ (blue) on left axis, $\Lambda$ (black) on right axis. Colored dashed lines in (d) indicate again where the cuts in Fig.~\ref{fig:fig2}(b) are taken.}
	\label{fig:fig3}
\end{figure}

We proceed to study a case example where the evolution of $\Lambda$ as a function of Zeeman field can be used to distinguish a trivial ABS (with MBS-resembling conductance features) from a MBS. We consider the setup shown in Fig.~\ref{fig:fig3}(a) where a section of the MI is uncovered from the superconducting shell. This situation could be relevant for the setup discussed above, since there indeed is a small section of the nanowire next to the tunnel barrier gate which is not covered by the superconducting shell. This section can also have a local potential that is different from the proximitized region. This setup was considered in~\cite{Chiu2017,Moore2017} and has
been shown to host trivial ABSs that stick to zero energy in a way that resembles MBSs. We found similar behavior for a $1.15$~\si{\micro\meter} long nanowire, with an uncovered section of length $L_N=0.15$~\si{\micro\meter}. We set the chemical potential $\mu=1.5$~\si{\milli\electronvolt}, so the nanowire is topologically trivial in the whole range of Zeeman field considered. The local potential is given by $V(x) = \Theta(L_N-x)V_0$, where $\Theta$ is the Heavyside step fuction and $V_0=1$~\si{\milli\electronvolt}. The remaining parameters are identical to those given in the previous sections.

Another widely used configuration with Majorana nanowires, for tunneling spectroscopy, is having a tunnel barrier only at one end of the wire, while the other end is open such that the MI is grounded and there are no charging effects. We calculate the finite temperature differential conductance in this configuration for the setup of the previous section, hosting MBSs, and the case example, hosting trivial ABSs. As seen by comparing Figs.~\ref{fig:fig3}(b,c), trivial ABSs might give rise to conductance features that are in practice indistinguishable from those of a MBS. The finite-temperature differential conductance is calculated by a convolution of the derivative of the Fermi function $n_F'(\epsilon)$ and the zero-temperature differential conductance $G_0$, which is calculated using a scattering-matrix formalism, described e.g.~in~\cite{Hansen2016}: $G_T(V_{\rm sd}) = -\int d\epsilon\: G_0(\epsilon)n_F'(\epsilon-V_{\rm sd})$.

Focusing on the same two systems, but now in a Coulomb-blockaded configuration, we plot their $\pm E_0$ and $\Lambda$ as a function of $V_{\rm Z}$ in Figs.~\ref{fig:fig3}(d,e). Comparing Fig.~\ref{fig:fig3}(e) with (d), we see that even though the oscillations of $E_0$ resemble MBS oscilations, the behavior of $\Lambda$ is qualitatively very different. That is, $\Lambda$ is not maximal/minimal when $E_0$ crosses zero and not $\approx 1/2$ when $|E_0|$ is maximal. In this way, by studying $\Lambda$ as a function of experimental parameters, such as the Zeeman field, a trivial ABS could be distinguished from a true MBS. However, it might be possible that a specific set of parameters could result in a topologically trivial ABS where both the evolution of $E_0$ and $\Lambda$ with Zeeman field resembles the MBS, but this will most likely be a very fine tuned situation.

\textit{Conclusion:} In this paper we have shown how the electron and hole components of the lowest-energy state in a Coulomb-blockaded superconducting island can be related to the measured
zero-bias conductance in the sequential-tunneling regime. In the case of the island hosting MBSs,
we found a characteristic way in which
the electron and hole components should oscillate as a function of magnetic field. This might be utilized to identify topologically trivial cases where a MBS-like zero bias peak is observed, by studying the magnetic-field dependence of the ratio $\Lambda$, as exemplified in the case example.

\section{Acknowledgments}
We gratefully acknowledge very helpful discussions with S.~M.~Albrecht and M.~Hell. This work was supported by the Danish National Research Foundation, by the DFG within the CRC 183 (project C01), and by the Research Council of Norway through its Centres of Excellence funding scheme, project number 262633, QuSpin.

\end{document}